\documentclass[showpacs,amsmath,
reprint,
aip]{revtex4-1}

\usepackage{graphicx}
\begin{document}

\title
{Stark effect of excitons in individual air-suspended carbon nanotubes}
\affiliation{Institute of Engineering Innovation, 
The University of Tokyo, Tokyo 113-8656, Japan}
\author{M.~Yoshida}
\author{Y.~Kumamoto}
\author{A.~Ishii}
\author{A.~Yokoyama}
\author{Y.~K.~Kato}
\email[Author to whom correspence should be addressed. Electronic mail: ]{ykato@sogo.t.u-tokyo.ac.jp}

\begin{abstract}
We investigate electric-field induced redshifts of photoluminescence from individual single-walled carbon nanotubes. The shifts scale quadratically with field, while measurements with different excitation powers and energies show that effects from heating and relaxation pathways are small. We attribute the shifts to the Stark effect, and characterize nanotubes with different chiralities. By taking into account exciton binding energies for air-suspended tubes, we find that theoretical predictions are in quantitative agreement.
\end{abstract}

\maketitle

Understanding of electric field effects on optical properties of single-walled carbon nanotubes (SWCNTs) is important for applications in nanoscale optoelectronic devices.\cite{Avouris:2008} One of the intriguing electro-optic effects is the Stark effect, which causes redshifts on exciton resonances under an application of electric fields.\cite{Perebeinos:2007, Zhao:2007} The effect has been used to explain spectral changes in electroabsorption,\cite{Takenobu:2006, Kishida:2008, Ogawa:2010} photoconductivity,\cite{Mohite:2008} and ultrafast measurements.\cite{Soavi:2013}  Local variation of excitonic energies \cite{Georgi:2010} and spectral diffusion at low temperatures \cite{Matsuda:2008prb2, Hogele:2008} have also been attributed to the Stark effect. These experiments, however, have been performed on ensembles of nanotubes,\cite{Takenobu:2006, Kishida:2008, Ogawa:2010} without well-defined electric fields,\cite{Matsuda:2008prb2, Hogele:2008, Georgi:2010, Soavi:2013} or on nanotubes with unknown chirality,\cite{Takenobu:2006, Mohite:2008} making quantitative analysis difficult.

Here we investigate field-induced redshifts of $E_{11}$ exciton emission in chirality-assigned individual SWCNTs. Photoluminescence (PL) spectra of air-suspended nanotubes within field-effect transistor structures are collected under an application of symmetric bias voltages on source and drain contacts, revealing redshifts that scale quadratically with electric field. We find that the shifts do not depend much on excitation power or energy, ruling out effects from heating or relaxation pathways. Attributing the redshifts to the Stark effect, we have also performed measurements on different chiralities, and a reasonable agreement with theoretical predictions \cite{Perebeinos:2007} is obtained by considering exciton binding energies for air-suspended tubes. Analysis using the total field rather than the longitudinal component shows more consistency, suggesting that transverse fields induce shifts of similar magnitude.

\begin{figure}
\includegraphics{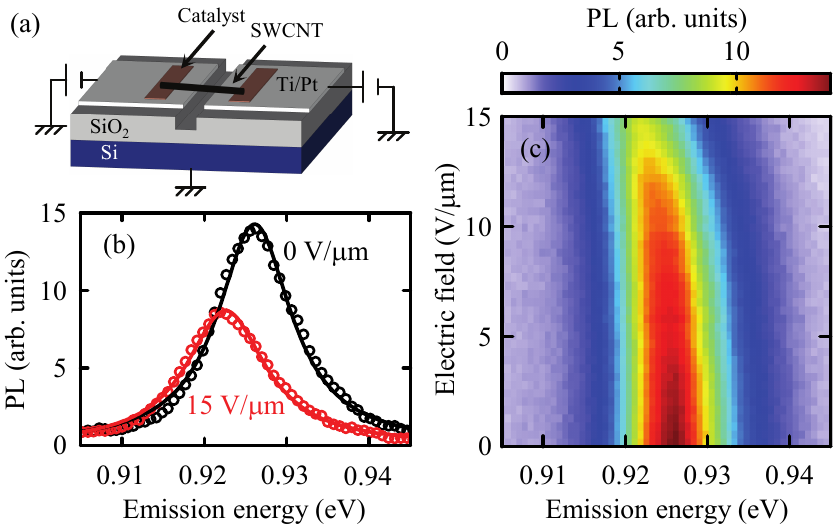}
\caption{\label{fig1}
(a)~A schematic of a device.
(b)~PL spectra taken with electric fields of 0~V/$\mu$m (black) and 15~V/$\mu$m (red). Open circles are data, and lines are Lorentzian fits.
(c)~PL spectra as a function of electric field. Data in (b-c) are taken with excitation energy of 1.651~eV and excitation power of 10~$\mu$W, and a (10,6) tube with a suspended length of 1.3~$\mu$m characterized in Ref.~\citenum{Kumamoto:2014} is used. Laser polarization is parallel to the nanotube axis.
}\end{figure}

Our field-effect transistors with suspended nanotubes \cite{Kumamoto:2014} are fabricated from Si substrates with 1-$\mu$m-thick oxide [Fig.~\ref{fig1}(a)]. Electron beam lithography is performed to pattern trenches with widths ranging from 1.0 to 1.7~$\mu$m, and an inductively coupled plasma etcher using CHF$_3$ gas is used to form 500-nm-deep trenches into the oxide layer. Another electron-beam lithography step defines the electrode patterns, and we deposit Ti (3~nm) and Pt (45~nm) using an electron beam evaporator with a base pressure of $1\times10^{-4}$~Pa. Following a lift-off process for the contacts, catalyst windows near the trenches are patterned with a third electron beam lithography step. Catalyst solution for nanotube growth is prepared by ultrasonicating 5~mg of cobalt(II) acetate tetrahydrate and 25~mg of fumed silica in 10.0~g of ethanol, and then we spin coat and lift off the catalyst. Finally, we perform chemical vapor deposition to synthesize single-walled carbon nanotubes.\cite{Maruyama:2002, Imamura:2013} The samples are placed in a quartz tube furnace and heated in dry air while the temperature is ramped up to 700$^\circ$C over 12~min. After pumping out the dry air, the quartz tube is filled by Ar with 3\% H$_2$, and the temperature in the furnace is elevated to 800$^\circ$C. Ethanol is introduced as a carbon source by bubbling the carrier gas for 10~min.

The devices are characterized with a confocal micro-PL system optimized for detection of nanotube emission.\cite{Moritsubo:2010, Watahiki:2012, Yokoyama:2014} An output of a wavelength tunable Ti:sapphire laser is focused onto the sample by an objective lens with a numerical aperture of 0.8 and a working distance of 3.4~mm. PL is collected through a confocal pinhole corresponding to an aperture with 5.4~$\mu$m   diameter at the sample image plane. An InGaAs photodiode array and a spectrometer with a spectral resolution of $\sim 1$~meV are used to detect PL from nanotubes. We utilize an automated three-dimensional translation stage with a travel range of $\pm$10~mm and a resolution of 50~nm to take PL images and to find the nanotube positions. PL excitation spectroscopy is performed to determine the chiralities of individual SWCNTs using tabulated data for air-suspended tubes.\cite{Ohno:2006prb} The nanotube length can be calculated from the width of the trench and the tube angle obtained from laser polarization dependence of PL.\cite{Moritsubo:2010}

Simultaneous PL and photocurrent (PC) measurements are performed in the presence of electric field using two dc source meters.\cite{Kumamoto:2014} We ground the Si substrate and apply symmetric bias voltages between the source and drain contacts in order to avoid PL quenching caused by carrier-induced Auger recombination.\cite{Freitag:2009, Yasukochi:2011} The longitudinal component of electric field $F$ is obtained by dividing the applied voltage with the tube length. We average the current with the excitation laser on the sample while a spectrum is taken, and then a background spectrum and a dark current are collected after blocking the laser with a shutter.  The PL spectrum is obtained by taking the difference of the two spectra. All measurements are done in air at room temperature.

Figure~\ref{fig1}(b) shows PL spectra taken with longitudinal electric fields $F= 0$ and 15~V/$\mu$m on a (10,6) nanotube under excitation at the $E_{22}$ resonance and a power $P=10$~$\mu$W. The emission peak with $F=15$~V/$\mu$m shows a redshift as well as a reduction of intensity compared to $F=0$~V/$\mu$m. To investigate such a behavior in detail, we have measured the field dependence of PL spectra from 0 to 15~V/$\mu$m [Fig.~\ref{fig1}(c)]. The peak position, width, and area are extracted by fitting the spectra with a Lorentzian function, and we obtain good fits as shown in Fig.~\ref{fig1}(b). Such fits have been performed for all of the spectra in Fig.~\ref{fig1}(c) in order to quantitatively characterize the field dependence. 

\begin{figure}
\includegraphics{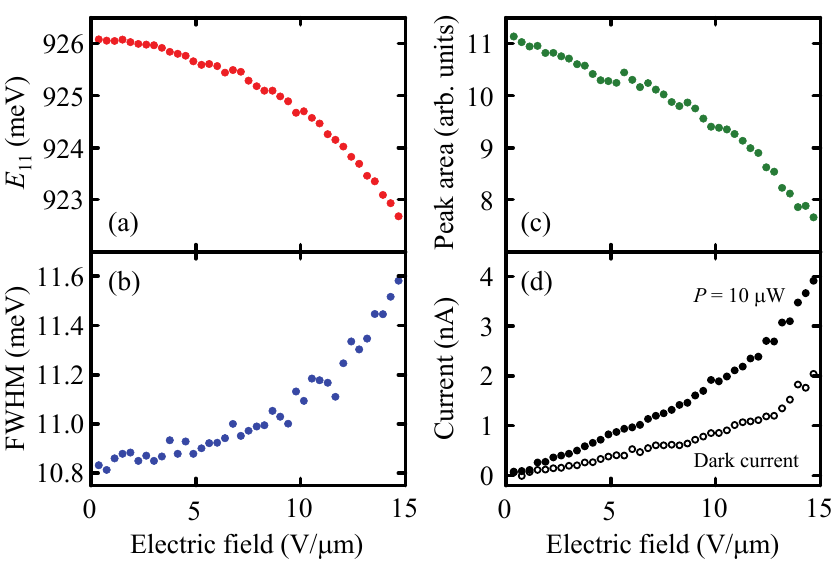}
\caption{\label{fig2}
(a-c)~Electric field dependence of emission energy, full width at half maximum (FWHM), and peak area, respectively, obtained from fits to data in Fig.~\ref{fig1}(c).
(d)~Current taken with $P=10$~$\mu$W (filled circles) and the dark current (open circles) as a function of electric field.
}\end{figure}

In Fig.~\ref{fig2}, we plot the electric field dependence of emission energy, linewidth, peak area, and current. The PL peak area decreases with increasing field as photocarriers flow into the contacts.\cite{Kumamoto:2014} We note that only those devices that show PC larger than 0.3~nA  are investigated here. If PC is small, it is likely that the nanotube has a bad contact to the electrodes, and the bias voltage can act as an effective gate. It is important to avoid such devices as unintentional electrostatic doping can occur, which is known to cause blueshifts.\cite{Yasukochi:2011}

The emission energy [Fig.~\ref{fig2}(a)] shows a quadratic redshift on the electric field, as expected from the Stark effect.\cite{Perebeinos:2007} The redshifts, however, could result from a rise in temperature\cite{Lefebvre:2004prb2} caused by Joule heating, as there exists a current flowing in the tube [Fig.~\ref{fig2}(d)]. Such a heating effect can be quantified from the linewidth broadening [Fig.~\ref{fig2}(b)]. In the case of  $F=15$~V/$\mu$m, the tube temperature increase is estimated to be $\sim$32~K by using a coefficient of 0.023~meV/K.\cite{Yoshikawa:2009}

We further consider whether the estimated rise in the temperature is reasonable. By solving a heat conduction equation in one dimension with a fixed temperature at the ends, we obtain a temperature increase of 2~K at the center of the tube for $F=15$~V/$\mu$m, assuming uniform Joule heating and a thermal conductivity of 3600~W/(K~m).\cite{Pop:2005} A contact thermal resistance of 490~K~m/W can explain the remainder of the temperature increase for a contact length of 1~$\mu$m, and this value is within the range observed for multi-walled nanotubes on Pt.\cite{Pettes:2009} Considering that the temperature estimate from the linewidth is justifiable, the field-induced redshift of more than 3~meV is difficult to explain by heating as we only expect 1~meV or smaller redshifts for such a temperature increase.\cite{Lefebvre:2004prb2}

\begin{figure}
\includegraphics{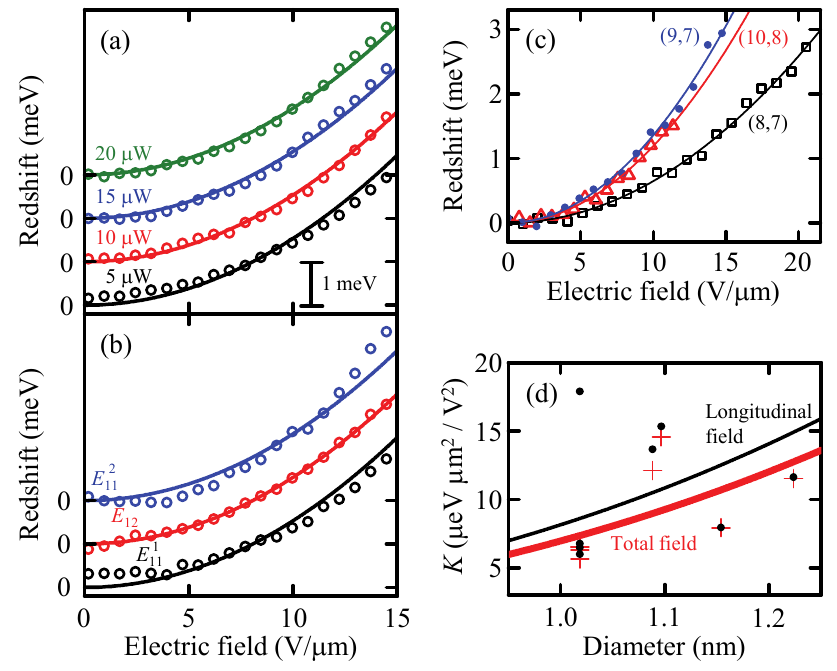}
\caption{\label{fig3}
(a)~Electric field dependence of $\delta E_{11}$ for different excitation powers. Data from bottom to top correspond to $P=5$, 10, 15, and 20~$\mu$W. The excitation energy is fixed at 1.651~eV. 
(b)~Redshifts as a function of $F$ taken at different excitation energies. Data from bottom to top are taken at $E_{11}^1$ (1.351~eV), $E_{12}$ (1.476~eV), and $E_{11}^2$ (1.450~eV) resonances,\cite{Lefebvre:2008} where $P=150$, 1000, and 200~$\mu$W are used, respectively.
For (a-b), the tube is the same as in Fig.~\ref{fig1} and Fig.~\ref{fig2}, and the data are offset for clarity.
(c)~Stark effect for a 1.0-$\mu$m-long (8,7) tube (squares), a 1.3-$\mu$m-long (10,8) tube (triangles), and a 1.0-$\mu$m-long (9,7) tube (dots), measured with $E_{22}$ excitations. For the (8,7), (10,8), and (9,7) tubes, $P=8$, 10, and 5~$\mu$W are used, and the redshifts are measured from 998.5, 862.7, and 954.7~meV, respectively. 
(d)~Diameter dependence of $K$. Dots and crosses show data analyzed with longitudinal field and total field, respectively.
The laser polarization is parallel to the tubes except for $E_{12}$ excitation in (b) where it is perpendicular. Lines are fits as explained in the text.
}\end{figure}

In order to directly rule out the heating-induced redshifts, we have investigated the electric-field dependence of the emission energy at different excitation powers [Fig.~\ref{fig3}(a)]. Because PC increases proportional to the excitation power, \cite{Kumamoto:2014} the Joule heating changes by a factor of four for $P=5$~$\mu$W and 20~$\mu$W. Any temperature-induced effects should show up as power dependent shifts, but we do not observe much difference, confirming that the thermal effects are limited.

Measurements at different excitation energies have also been performed [Fig.~\ref{fig3}(b)], and the electric-field dependence shows similar quadratic behaviors for excitation at the $E_{12}$, $E_{11}^1$, and $E_{11}^2$ states.\cite{Lefebvre:2008} These results show that relaxation pathways from the excited states to the $E_{11}$ state do not influence the shifts of the emission energy, and that the redshifts are the intrinsic response of the $E_{11}$ excitons under electric fields. 

We therefore attribute the redshift to the Stark effect, and to characterize the strength of the effect, the data are fit to
\begin{equation}
\delta E_{11} = KF^2,\label{eq1}
\end{equation}
where $\delta E_{11}$ is the redshift and $K$ is a constant. Reasonable fits are obtained as shown in Fig.~\ref{fig3}(a) and \ref{fig3}(b) using a single value of $K=15.3$~$\mu$eV~$\mu$m$^2/$V$^2$. 

Nanotubes with different chiralities have also been investigated [Fig.~\ref{fig3}(c)]. All of the tubes show redshifts that increase quadratically with electric fields, but the amounts of the shifts are different. The values of $K$ obtained for the eight tubes measured are plotted as a function of the tube diameter $d$ [dots in Fig.~\ref{fig3}(d)]. The diameter dependence of the Stark effect is expected to be \cite{Perebeinos:2007}
\begin{equation}
K= \kappa_\text{b}\frac{(ed)^2}{E_\text{b}}, \label{eq2}
\end{equation}
where $\kappa_\text{b}$ is a unitless constant, $e$ is the electronic charge, and $E_\text{b}=0.66/d$~eV is the exciton binding energy for air-suspended nanotubes. \cite{Lefebvre:2008} The thin line in Fig.~\ref{fig3}(d) shows the fit to the data using Eq.~\ref{eq2}. We obtain $\kappa_\text{b}=5.4$, which is comparable to the theoretical values ranging from 3.0 to 4.2 in Ref.~\citenum{Perebeinos:2007}, although the data has a large scatter. 

In particular, of the four (8,7) nanotubes we have measured, one shows a relatively large $K$. It turns out that this nanotube is suspended at an angle of $37^{\circ}$ with respect to the trench. Since the electric field is established between the source and drain contacts, this tube is at a large angle with the field. If transverse fields can also induce Stark shifts,\cite{Takenobu:2006, Mohite:2008} it may explain the deviation as the analysis so far only considered the longitudinal component.

When we replot the data using the total field [crosses in Fig.~\ref{fig3}(d)] as opposed to the longitudinal field, we find that $K$ of the four (8,7) tubes falls within $\pm$9\%. In addition, $\kappa_\text{b}=4.6$ is obtained [thick line in Fig.~\ref{fig3}(d)], which is closer to the theoretical values. Although the data still show some deviations from Eq.~\ref{eq2}, our results are in reasonable agreement as the calculations also find chilarity dependent $\kappa_\text{b}$ that can differ by $\pm$17\%.

With the reasonable agreement with theory, we can estimate the exciton dissociation time under the experimental conditions. The dissociation rate can be written as \cite{Perebeinos:2007}
\begin{equation}
\Gamma = \alpha E_\text{b} \frac{F_0}{F}\exp\left(-\frac{F_0}{F}\right), \label{eq3}
\end{equation}
where $\alpha=4.1$ is a constant and $F_0$ is the dissociation field. For a $d=1.1$~nm tube with $E_\text{b}=288$~meV, $F_0$ has been calculated to be 198~V/$\mu$m in Ref.~\citenum{Perebeinos:2007}. Using the relation $F_0\propto E_\text{b}^{3/2}$, we estimate $F_0=595$~V/$\mu$m for an air-suspended (10,6) tube by scaling for $E_\text{b}=600$~meV. The dissociation time $\hbar/\Gamma$, where $\hbar$ is the Planck constant, is longer than 1~s for the highest applied field of 15~V/$\mu$m, and as the lifetime of $E_{11}$ excitons is much less than 1~ns,\cite{Xiao:2010} we do not expect $E_{11}$ exciton dissociation at these fields.\cite{Kumamoto:2014}

In summary, we have investigated the redshifts of PL emission from individual air-suspended SWCNTs caused by electric fields. Since the shifts are quadratic in field and not dependent on excitation power or energy, they are attributed to the Stark effect. By performing measurements on different chiralities, we have shown that the shifts are in quantitative agreement with theoretical calculations. The results on tubes at large angles with the field suggest that transverse fields may be causing Stark shifts of similar magnitude. As the estimated dissociation rates are negligible, it should be possible to achieve wider control of emission wavelength in stronger fields or larger diameter tubes.

\begin{acknowledgments}
We thank T. Kan and I. Shimoyama for the use of the evaporator, and S. Yamamoto for the plasma etcher. This work is supported by KAKENHI (24340066, 24654084, 26610080), the Canon Foundation, Asahi Glass Foundation, and KDDI Foundation, as well as the Photon Frontier Network Program of MEXT, Japan. The devices were fabricated at the Center for Nano Lithography \& Analysis at The University of Tokyo.
\end{acknowledgments}

\end{document}